\documentclass[pre,twocolumn,amsmath,amssymb,showpacs]{revtex4}  
\usepackage{wasysym}
\usepackage{epsf}
\usepackage{graphicx}
\begin{document}
\title{Contactless inductive flow tomography}
\author{Frank Stefani, Thomas Gundrum,  and Gunter Gerbeth}
\affiliation{Forschungszentrum Rossendorf\\
P.O. Box 510119, D-01314 Dresden, Germany}

\date{\today}

\begin{abstract}
The three-dimensional velocity field
of a propeller driven liquid metal flow
is reconstructed
by a contactless inductive flow tomography (CIFT).
The underlying theory  is presented within the framework of
an integral equation system that governs the magnetic field
distribution in a moving electrically conducting fluid.
For small magnetic Reynolds numbers this integral equation system
can be cast into a linear inverse problem for the determination
of the velocity field from externally measured magnetic fields.
A robust reconstruction of the large scale velocity field is already 
achieved
by applying the external magnetic field alternately in two orthogonal
directions and measuring the corresponding sets of induced magnetic fields.
Kelvin's theorem is exploited to regularize the resulting velocity field
by using the kinetic energy of the flow as a
regularizing functional. 
The results of the new  technique are shown to be in           
satisfactory agreement with ultrasonic measurements.                     
\end{abstract}


\maketitle

\section{Introduction}                                                     

Flow measurement in metallic and semiconducting melts is a
notorious problem in a number of technologies, reaching
from iron casting to silicon
crystal growth. 
Obviously, the usual optical methods 
of flow measurement
are inappropriate for those opaque fluids. Ultrasonic techniques have 
problems, too,
when applied to very hot or chemically aggressive melts. A completely 
contactless flow measurement technique would be highly desirable, even 
if it were only to provide a rough picture of the flow. 

Fortunately, metallic and semiconducting melts are characterized by a high
electrical conductivity. Hence, when exposed to an external magnetic field, 
the flowing melt gives rise to electrical currents that lead to a 
deformation of the applied
magnetic field. This field deformation is measurable outside the
fluid volume, and it can be used to 
reconstruct the velocity field, quite in parallel with 
the well-known
magnetoencephalography, where
neuronal activity in the brain is inferred 
from  magnetic field measurements \cite{HAEM}. The 
goal of this paper is to report on a first experimental demonstration 
of such a {\it contactless inductive flow tomography (CIFT)}.

\section{Theory}                                                           

The ratio of the induced field to the applied field is 
determined by the 
so-called magnetic Reynolds number, defined as $Rm=\mu \sigma v l$, with 
$\mu$ denoting the magnetic permeability of the melt, 
$\sigma$ its electrical conductivity, $v$ a typical velocity, and $l$ a typical 
length scale of the flow. In industrial applications, $Rm$ is in the order
of 0.01...1. Only for a few large scale sodium flows, as they appear in fast
breeder 
reactors, but also in the recent hydromagnetic dynamo experiments \cite{RMP},  
$Rm$ reaches values
in the order of 10...100 (of course, in some cosmic dynamos $Rm$ can be
even much larger). Actually, the present work was strongly motivated by 
the wish to reconstruct the sodium flow in the Riga dynamo experiment by
an appropriate contactless method. 

Suppose the fluid to flow with the stationary velocity $\bf v$,  
and to be exposed to a magnetic field ${\bf B}$, which we leave unspecified 
for the  moment. 
Then, according to Ohms law in moving conductors 
the current
\begin{eqnarray}
\bf{j}=\sigma (\bf{v} \times {\bf{B}}-\nabla \varphi)
\label{eq1}
\end{eqnarray}
is induced, with $\varphi$ denoting the electric potential. This current
gives rise to the induced magnetic field
\begin{eqnarray}
{\bf {b}}({\bf {r}})&=&\frac{\mu_0 \sigma}{4 \pi} \iiint_V
\frac{ ({\bf {v}}  ({\bf {r'}}) \times {\bf {B}}  ({\bf {r'}}))  \times 
({\bf {r}}-{\bf {r'}})}{|{\bf {r}}-{\bf {r'}}|^3} \; 
dV' \nonumber \\
&&-\frac{\mu_0 \sigma}{4 \pi} \oiint_S \varphi({\bf {s'}}) 
{\bf {n}} ({\bf {s'}}) \times
\frac{{\bf {r}}-{\bf {s'}}}{|{\bf {r}}-{\bf {s'}}|^3} 
\; dS' \; .
\label{eq2}
\end{eqnarray}
Equation (\ref{eq2}) follows from inserting Eq. (\ref{eq1})
into  Biot-Savart's law and transforming the volume integral over $\nabla \varphi$ 
into a surface integral over $\varphi$. 

The electric potential $\varphi$ at the boundary $S$, in turn, 
has to fulfill the boundary integral equation 
\begin{eqnarray}
\varphi({\bf {s}})&=& 
\frac{1}{2 \pi} \iiint\limits_D  
\frac{ ({\bf {v}}  ({\bf {r'}}) \times {\bf {B}}  ({\bf {r'}})) 
\cdot ({\bf {s}}-{\bf {r'}})}{|{\bf {s}}-
{\bf {r'}}|^3} 
\; dV' \nonumber \\
&& -\frac{1}{2 \pi} \oiint\limits_S \varphi({\bf {s'}})
{\bf {n}}({\bf {s'}}) 
\cdot \frac{{\bf {s}}-{\bf {s'}}}{{|{\bf {s}}-{\bf {s'}}|}^3} 
\;  dS' \; .
\label{eq3}
\end{eqnarray}
Equation (\ref{eq3})
follows from taking the divergence of Eq. (\ref{eq1}) and utilizing
$\nabla \cdot {\bf j}=0$. Then, Green's theorem can be applied to the solution of the
arising Poisson equation  $\Delta \varphi =\nabla \cdot ({\bf {v}} \times  {\bf {B}})$,
with demanding that the current is purely tangential at the boundary \cite{INV1}.
Note that Eq. (\ref{eq3}) is the basic formula
for the vast area of {\it electric} inductive flow measurement
\cite{SHER} which is, however, not the subject of the present work.

In general, the magnetic field $\bf B$ on the right hand sides of Eqs. 
(\ref{eq1}-\ref{eq3})
is the sum of an externally applied magnetic field ${\bf B}_0$ and the very
induced magnetic field $\bf b$. Hence 
Eqs. (\ref{eq2},\ref{eq3}) represent 
an integral
equation system which actually can be used to solve dynamo problems
in arbitrary bounded domains  \cite{XU}.
It it also suitable for a systematic investigation of the
non-linear induction effects
as they appear already in the sub-critical regime of laboratory
dynamos \cite{VKS}. 

\begin{figure}[t]
\epsfxsize=8.6cm\epsfbox{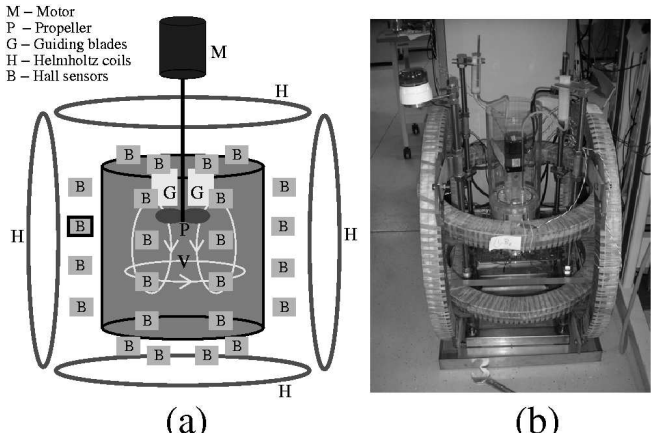}
\caption{Scheme (a) and photograph (b) of the CIFT experiment.}
\end{figure}
 
In the following, however, all considerations will be restricted to 
problems with  small $Rm$ for which $\bf B$ can be replaced by
${\bf B}_0$. Then, we  get a linear relation between the 
desired velocity field 
and the induced magnetic field which is supposed to be measured.
But how to cope with the remaining Eq. (\ref{eq3}) for the
electric potential? 

\begin{figure}[t]
\epsfxsize=8.6cm\epsfbox{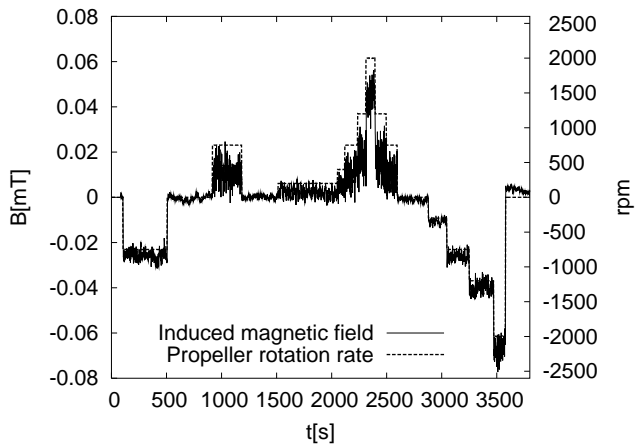}
\caption{Propeller rotation rate, and induced magnetic field measured at 
the Hall sensor emphasized in Fig. 1a.}
\end{figure}

The answer to this question can be adopted from
magnetoencephalography \cite{HAEM}.  
Assume, for a given ${\bf B}_0$, all measured magnetic field data be collected 
into an $N_B$ dimensional vector with the entries $b^{(B_0)}_i$, and the desired 
velocity components at the $N_V$ discretization points by a vector with the
entries $v_n$. The solution of the boundary integral equation may require
a fine discretization of the boundary, with $N_P$ degrees of freedom
$\varphi^{(B_0)}_k$.
Eqs. (\ref{eq2},\ref{eq3}) can then be written in the form
\begin{eqnarray}
b^{(B_0)}_i&=&R^{(B_0)}_{in} v_n+S_{ik} \varphi^{(B_0)}_k \label{eq4} \\
\varphi^{(B_0)}_k&=&T^{(B_0)}_{kn} v_n+U_{kk'} \varphi^{(B_0)}_{k'} \; ,\label{eq5}
\end{eqnarray}
where the matrices ${\bf{R}}^{(B_0)}$ and ${\bf{T}}^{(B_0)}$ depend on the 
applied field ${\bf B}_0$, whereas the matrices $\bf S$ and $\bf U$ depend only
on geometric factors.

\begin{figure}[t]
\epsfxsize=8.6cm\epsfbox{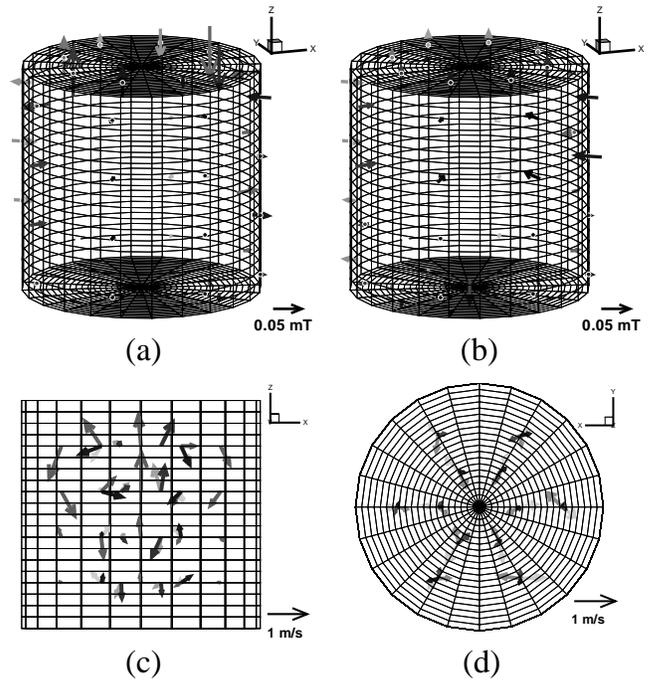}
\caption{Measured induced magnetic field components for transverse (a) and axial (b)
applied magnetic field, and  reconstructed velocity as seen 
from the side (c) and from below (d).
The grey scale of the arrows indicates the  distance from the eye.
The propeller pumps {\it upward} with 1200 rpm.}
\end{figure}

\begin{figure}[t]     
\epsfxsize=8.6cm\epsfbox{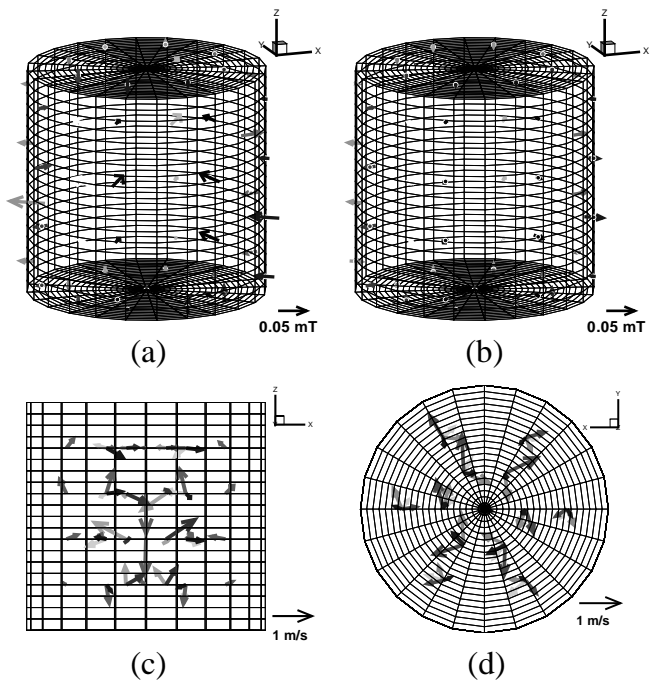}
\caption{The same as Fig. 3, but for the propeller pumping {\it downward} 
with  1200 rpm.}
\end{figure}

As is well known from magnetoencephalography,                               
the inversion of Eq. (\ref{eq5}) is a bit tricky due to the singularity 
of the matrix $({\bf {I}}-{\bf{U}})$. This singularity mirrors the fact 
that the electric potential is defined only up to an additive constant.
We can remove this ambiguity by replacing $({\bf {I}}-{\bf{U}})$  
by a generally well
conditioned matrix 
$({\bf {I}}-{\bf{U}})^{defl}:=({\bf {I}}-{\bf{U}})-N^{-1} {\bf e} {\bf e}^T$, 
where ${\bf e}$ is a vector with all 
$N$ entries equal to one and ${\bf e}^T$ is its transposed.
By applying this so-called deflation method 
\cite{HAEM}
one ends up with
\begin{eqnarray}
b^{(B_0)}_i&=&R^{(B_0)}_{in} v_n+S_{ik'}(I-U)^{-1,defl}_{k'k} 
T^{(B_0)}_{kn} v_n \; ,\label{eq6} 
\end{eqnarray}
i.e., with a linear relation between the desired velocity field and the 
measured magnetic field.

\begin{figure}[t]
\epsfxsize=8.6cm\epsfbox{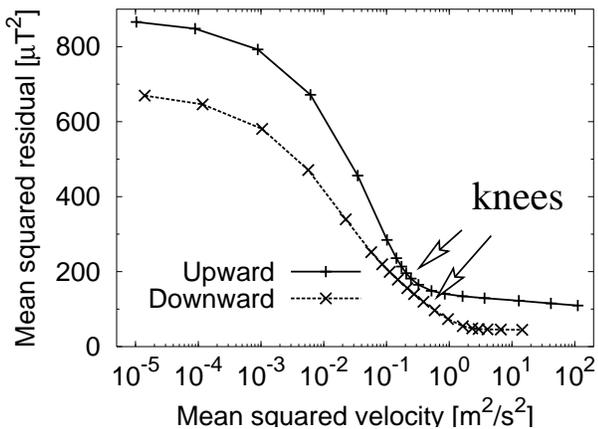}
\caption{Tikhonov's L-curve for the two different pumping directions.
The arrows point
at the bended knee where the curves have the strongest curvature.
At these points we get a reasonable compromise between data fitting and
minimum kinetic energy of the modelled velocity field.
The rms of these velocities is approximately 0.41 m/s for the
upward pumping and 0.73 m/s for the downward pumping.}
\end{figure}

Despite the far-reaching similarity, there is one essential difference 
of our method 
compared to magnetoencephalography. While in the latter one has to determine
a single
neuronal current distribution, in our case  we 
can produce quite different current distributions {\it from the same flow field} 
simply by 
applying various external magnetic fields subsequently. For each
applied magnetic field 
we can measure the corresponding induced fields, and utilize all 
of them to reconstruct the flow.

Concerning the uniqueness question for this sort of inversion, here we 
give only a shortened answer, referring for more details to the previous 
papers \cite{INV2,MEAS}. 
For spherical geometry, and the two applied magnetic fields pointing in
orthogonal directions, the problem can be solved with some 
rigour.
Suppose we have measured the two corresponding sets of induced 
magnetic fields on a sphere outside the fluid volume, and have 
expanded them into 
spherical harmonics. The desired (solenoidal) velocity field can be 
represented by two scalars
for its poloidal and toroidal parts. These two scalars can also be expanded
into spherical harmonics, but with the expansion coefficient still being
functions of the radius. In \cite{MEAS} it had 
been shown (at least in some low degrees of the
spherical harmonics expansion) that what can be derived from the
two magnetic field expansion coefficients are some radial moments of the
expansion coefficients for the velocity field.
A further concretization of the radial dependence of the velocity
expansion coefficients can only be achieved by regularization techniques. 
If we demand, in a slight overinterpretation of Kelvin's theorem, 
the flow to possess 
minimal kinetic energy, we obtain 
a unique solution 
for the radial dependence, too. 

Without any rigorous proof at hand, we assume that this result
can  be generalized to aspherical geometry: the large main structure of the 
large scale flow 
is well inferable, with a depth ambiguity of the velocity that can only be 
resolved by regularization techniques.
Imposing two orthogonal magnetic fields represents a certain minimum
configuration for such a flow tomography. 
For a single magnetic field of one direction there are, of course, 
flow components which would be hidden from outside. However, all those 
components are 
detectable for an external magnetic field orthogonal to the previous one.

For our experimental application  we   
employ the so-called Tikhonov regularization \cite{HANS},  
minimizing the total functional
\begin{eqnarray}
F[{\bf{v}}]=F_{B_{0x}}[{\bf{v}}]+F_{B_{0z}}[{\bf{v}}]+
F_{div}[{\bf{v}}]+F_{reg}[{\bf{v}}]
\label{eq7}
\end{eqnarray}
with
\begin{eqnarray}
F_{B_{0x}}[{\bf{v}}]&=&\sum_{i=1}^{N_B}\frac{1}{\sigma^2_{i}}
\left( b_{i,meas}^{(B_{0x})}-b_{i}^{(B_{0x})}[{\bf{v}}]\right)^2
\label{eq8}\\
F_{B_{0z}}[{\bf{v}}]&=&\sum_{i=1}^{N_B}\frac{1}{\sigma^2_{i}}
\left( b_{i,meas}^{(B_{0z})}-b_{i}^{(B_{0z})}[{\bf{v}}]\right)^2
\label{eq9}\\
F_{div}[{\bf{v}}]&=&\frac{1}{\sigma^2_{div}}
\sum_{k=1}^{N_V}\left(\nabla \cdot {\bf{v}}\right)^2_k \Delta V_k
\label{eq10}\\
F_{reg}[{\bf{v}}]&=&\frac{1}{\sigma^2_{pen}}
\sum_{k=1}^{N_V}{\bf{v}}^2_k \Delta V_k \; .
\label{eq11}
\end{eqnarray}
The first two functionals represent, for applied transverse field ${\bf B}_{0x}$
and axial field  ${\bf B}_{0z}$, respectively, the mean 
squared residual deviation of the measured induced 
magnetic fields $b_{i,meas}^{(B_{0})}$ from the 
fields $b_{i}^{(B_{0})}[{\bf v}]$ modeled according to Eq. (\ref{eq6}). 
$F_{div}[{\bf{v}}]$ 
enforces the velocity field to be solenoidal,
and $F_{reg}[{\bf{v}}]$ is the regularization functional which 
tries to minimize the kinetic energy. 
The parameters $\sigma_{i}$ are the assumed 
a-priori errors for the measurement of the induced fields. 
The parameter $\sigma_{div}$ is chosen very small as it
is a measure 
for the
divergence the velocity solution is allowed to have.  
The parameter $\sigma_{pen}$ determines the trade-off between minimizing the
mean squared residual deviation of the observed fields and minimizing the
kinetic energy of the estimated velocity field. 
The normal equations, that follow from the minimization of the functional 
(\ref{eq7}), 
are solved by Cholesky decomposition.

\section{Experiment}                                                  

In the experiment (Fig. 1)
we use 4.4 liters of the 
eutectic alloy Ga$^{67}$In$^{20.5}$Sn$^{12.5}$ 
that is liquid at room temperatures. 
The flow is produced by a motor driven propeller
with a diameter of 6 cm inside a cylindrical 
polypropylene vessel with 18.0 cm 
diameter. The height of the liquid metal is 17.2 cm, yielding
an aspect ratio close to 1.

\begin{figure}[t]     
\epsfxsize=8.6cm\epsfbox{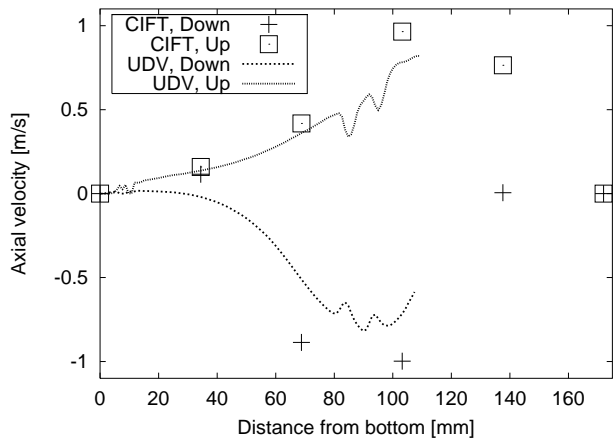}
\caption{Axial velocities along the central vertical axis of the the 
cylinder, determined by CIFT and by ultrasonic measurements (UDV).
The ultrasonic measurements are only shown up to the
propeller position, whereafter they become unreliable.}
\end{figure}

The position of the propeller is 
approximately at one
third of the total hight, measured from the top. 
Eight guiding blades above the propeller are intended to remove 
the swirl of the flow for the case that the propeller pumps upward.   
Contrary to that, the downward pumping produces, in addition to the main
poloidal motion, a considerable toroidal motion. The rotation rate of
the propeller can reach up to 2000 rpm, which amounts to a 
mean velocity of approximately 1
m/s, corresponding to a magnetic Reynolds number of
approximately 0.4.

\begin{figure}[t]    
\epsfxsize=6.6cm\epsfbox{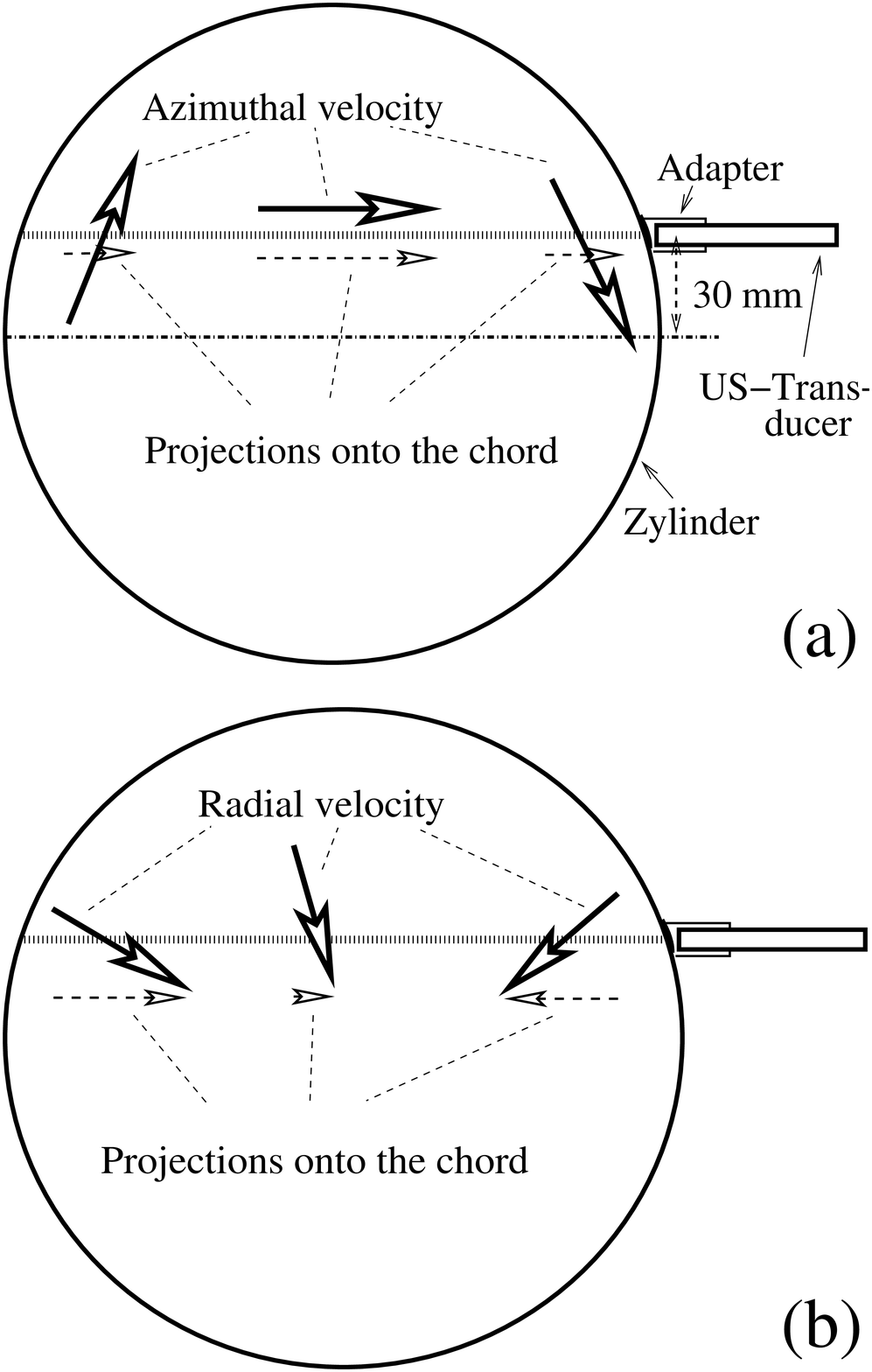}
\caption{Sketch of the ultrasonic measurement set-up for the azimuthal 
velocity component. The obtained velocity along the chord is 
a position dependent mixture of radial and azimuthal components. In the middle
of the chord, one gets the pure azimuthal component. 
(a) Typical situation for downward pumping with dominant azimuthal velocity
(at the axial position 70 mm from the bottom). The
projection of the velocity onto the chord has a maximum in the middle of the chord.
(b)  Typical situation for upward pumping with dominant radial velocity.
The projection of the velocity goes to zero in the middle of the chord.}
\end{figure}

Two pairs of Helmholtz coils are fed by currents of 22.5 Ampere 
and 32.5 Ampere, respectively, to produce alternately an axial and a 
transversal field of
4 mT, which both are rather homogeneous throughout the vessel.
Either field is
applied for a period of 3 seconds, during which a trapezoidal
signal form is used.
The measurements are carried out for 0.5 seconds, 1 second after the 
plateau value of the trapezoidal current has been reached.
Hence, we get an online monitoring with a time resolution
of 6 seconds.

\begin{figure}[t]     
\epsfxsize=8.6cm\epsfbox{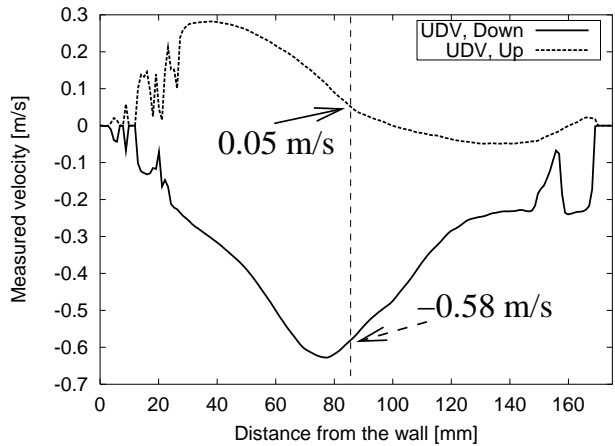}
\caption{Measured projection of the velocity onto the chord, for 
downward and upward pumping. In the middle of the chord (at 85 mm distance 
from the wall) we get an azimuthal velocity of +0.58 m/s for downward pumping 
and -0.05 m/s for upward pumping (note the change of sign due to conventions 
in UDV measurements). Compare also Fig. 7 for illustration.}
\end{figure}

The induced magnetic fields are measured by 49 Hall sensors,
8 of them grouped together on each of 6 circuit boards which
are located on different heights (Fig. 1). One additional
sensor is located in the center below the vessel. 
The key problem of the  method
is the reliable determination of comparably 
small induced magnetic fields
on the background of much higher imposed magnetic fields.
An accurate control of the
external magnetic field is essential to meet this goal. 
In our configuration the current drift
 in the Helmholtz coils can be controlled with an accuracy of
 better than 0.1 per cent. This is sufficient since the measured
induced fields are approximately 1 per cent of the applied field.
The temperature drift of the sensitivity  can
be overcome by enforcing the applied  current in the Hall sensors to be 
constant. The temperature drift of the offset problem is  circumvented
by changing the sign of the applied magnetic field. 
Figure 2 shows that by these means a stable measurement of the small 
induced field
can be realized, even over a period of one hour.

\begin{figure}[t]     
\epsfxsize=8.6cm\epsfbox{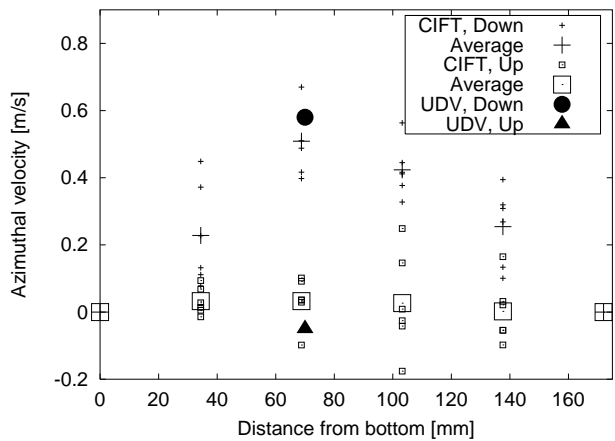}
\caption{Azimuthal velocity at r=30 mm, as determined by CIFT at different 
axial positions. The small symbols (crosses and squares) represent  individual
measurements at six azimuthal positions. 
The large crosses and squares represent the
corresponding averages over six azimuthal positions. The full symbols (circle and
triangle) are the UDV values as inferred from Fig. 8.}
\end{figure}

For upward and downward pumping, Figs. 3 and 4 show the 
induced magnetic fields measured at the 49 positions, 
and the inferred velocity field at 52 discretization points. In Fig. 3c
we see clearly the upward flow in the center of the vessel and 
the downward flow at the rim, but nearly no rotation of the
flow in Fig. 3d. In Fig. 4c we can identify the downward flow in the center and
the
upward flow at the rim, and in Fig. 4d a clear
rotation of the flow. Evidently, the method is able to identify 
the poloidal rolls and the 
absence or presence of the swirl.

For both flow directions, Fig. 5 illustrates the application of 
Tikhonov's L-curves  \cite{HANS}.
This curve, which  results from scaling the parameter $\sigma_{pen}$  in Eq. (11)
from lower to higher values, shows the dependence of the mean squared residual of the
measured data on the kinetic energy of the flow.
For low values (left end of Fig. 5) only little
kinetic energy is allowed, leading to a velocity field that  fits the 
measured magnetic field data only poorly. For high values (right end of Fig.5)
the data are fitted very well but with an unphysical high kinetic energy.
At the points of strongest bending (the ''knee''), the resulting velocities 
(Figs. 3 and 4)
are physically most reasonable \cite{HANS}.

\section{Validation}                                                   

In order to validate the CIFT method, we have performed independent
velocity measurements based on ultrasonic Doppler velocimetry (UDV).
For that purpose we have used the DOP2000 ultrasonic velocimeter 
manufactured by Signal-Processing SA (Lausanne, Switzerland), which had 
already demonstrated its capabilities for velocity
measurements in liquid metals \cite{ECKE,CRAMER}.
As ultrasonic
transducers we have used 2 MHz probes.

Because of its comparably large magnetic Reynolds number ($Rm \approx 0.2$), 
the propeller driven flow in the cylinder has also a large hydrodynamic Reynolds 
number 
($Re \approx 2 \times 10^5$). Necessarily, the flow is highly turbulent. 
Strong fluctuations are observed both by the CIFT method as well as 
by UDV.

For a sensible comparison of both methods, some time averaging is advised.
In the following we will focus on
two UDV measurements that were both taken at a propeller rotation rate of 
1200 rpm, and which represent a time average over half a minute.

The first measurement concerns the axial velocity along the central 
vertical axis of the cylinder. 
This axial component is easily measured 
by an ultrasonic transducer flash mounted to the bottom of the cylinder. 
Figure 6 shows the results of the UDV measurement (up to the propeller position),
together with the results of the CIFT measurement. For both upward and downward 
pumping we see a reasonable correspondence of both measurements. Notably, CIFT 
exhibits the different axial dependencies that are typical for upward and downward 
pumping and which are confirmed by the UDV data.

The second measurement, which concerns the azimuthal velocity component, deserves
some explanation. Figure 7 shows the UDV measurement set-up. The axial position is 
at 70 mm from the bottom.   What is actually measured by UDV is
the projection of the velocity onto the ultrasound beam along the chord. Therefore, 
the measured 
signal is in general a mixture of the radial and azimuthal velocity components. 
Only in  the middle of the 
chord we
get a signal that originates purely from the azimuthal velocity.
In Figs. 7 (a) and (b) we illustrate  the measured data that are shown in Fig. 8.
In the case of downward pumping  the velocity is dominated by the rotation 
whereas for upward
pumping it is dominated by the radial part. 
In the middle of the chord we infer a mean azimuthal velocity of 
0.58 m/s for downward pumping and -0.05 m/s for upward pumping.

Do these UDV values agree with those obtained by CIFT? In Fig. 9 we show
the axial dependence of the azimuthal velocity at a radial position r=30 mm, 
as obtained by
CIFT during a measurement time of 6 seconds. 
We give here the individual values at six different azimuthal positions 
(small crosses
and squares). Interestingly, though the system is in general axisymmetric, 
non-axisymmetric fluctuations are still visible. The averages (large crosses and 
squares) over the
six azimuthal positions show, at an 
axial position of z=70 mm, a good agreement with the
data from UDV measurement.

\section{Conclusions and prospects}
     
To summarize, we have put into practice a first version of contactless
inductive flow tomography, using  two orthogonal imposed magnetic fields. 
The comparison with UDV measurements shows that                                  
the method provides robust results on the 
main structure and the amplitude of the velocity field.
A particular power of CIFT consists in a transient resolution of the full 
three-dimensional flow structure in steps of several seconds. Hence, 
slowly changing flow fields in various processes can be followed in time.
Due to its weakness the externally applied magnetic field 
does not influence the flow to be measured. 
However, CIFT is also possible in cases where stronger magnetic fields are 
already present 
for the purpose of flow control, as, e.g., the electromagnetic brake in steel 
casting or the DC-field components in silicon crystal growth.
Obviously, the future of the method lays with applying
AC fields with different frequencies in order to improve the 
depth resolution
of the velocity field.
For problems with higher $Rm$, including dynamos, 
the inverse problem  becomes non-linear, and more
sophisticated inversion methods must be applied to infer the
velocity structure from magnetic field data. 
Although interesting results have been obtained by 
employing Evolutionary Strategies to 
inverse spectral dynamo 
problems \cite{OSZI}, and first tests of such inversion schemes for the data from 
the Riga dynamo             
experiment have shown promising results, the general inverse 
dynamo topic is extremely complicated and goes
essentially beyond the scope of the present paper.  

\section*{ACKNOWLEDGMENTS}

Financial support from German "Deutsche Forschungsgemeinschaft" 
under Grant No GE 682/10-1,2 is gratefully acknowledged.

\end{document}